\documentclass[11pt]{article}
\pdfoutput=1
\usepackage{amsmath,epsf,amssymb,latexsym,enumerate,cite,shadow,array,color}
\usepackage{slashed}
\usepackage{graphicx,wasysym}

\DeclareFontFamily{U}{rsf}{} \DeclareFontShape{U}{rsf}{m}{n}{
  <5> <6> rsfs5 <7> <8> <9> rsfs7 <10-> rsfs10}{}
\DeclareMathAlphabet\Scr{U}{rsf}{m}{n} \makeatletter
\def\be{\begin{equation}}
\def\ee{\end{equation}}
\def\ba{\begin{array}}
\def\ea{\end{array}}

\newcommand{\beq}{\begin{equation}}
\newcommand{\eeq}[1]{\label{#1}\end{equation}}
\newcommand{\bea}{\begin{eqnarray}}
\newcommand{\eea}[1]{\label{#1}\end{eqnarray}}

\def\Re{\mathop{\rm Re}\nolimits}
\def\Im{\mathop{\rm Im}\nolimits}

\usepackage{ifpdf}
\ifx\pdfoutput\undefined
   \pdffalse
\else
   \pdfoutput=1
   \pdftrue
  \usepackage[pdftex]{hyperref}
\begin{document}

\begin{titlepage}
\hfill CERN-TH-2017-164\\

\hskip 1.5cm

\begin{center}

{\huge \bf{

Comments on Nonlinear Sigma Models Coupled to Supergravity}
 }

\vskip 0.8cm  

{\bf \large Sergio Ferrara$^{a,b,c}$ and Massimo Porrati$^{d}$}  

\vskip 0.5cm

\noindent $^{a}$ Theoretical Physics Department, CERN, CH 1211, Geneva 23, Switzerland\\
$^{b}$ INFN - Laboratori Nazionali di Frascati, Via Enrico Fermi 40, I-00044 Frascati, Italy\\
$^{c}$ Department of Physics and Astronomy and Mani L. Bhaumik Institute for Theoretical Physics 
University of California Los Angeles, CA 90095-1547 USA\\
$^{d}$ {CCPP, Department of Physics, NYU,
726 Broadway, New York NY 10003, USA}

\end{center}

\vskip 1 cm

\begin{abstract}
 
$N=1$, $D=4$ non linear sigma models, parametrized by chiral superfields, usually describe K\"ahlerian 
geometries, provided that Einstein frame supergravity is used. 
The sigma model metric is no longer K\"ahler when 
local supersymmetry becomes nonlinearly realized through the nilpotency of the supergravity auxiliary fields.
In some cases the nonlinear realization eliminates one scalar propagating degree of freedom. This happens when the sigma model
conformal-frame metric has co-rank 2.  In the geometry of the inflaton, this effect eliminates its scalar superpartner. 
We show that the sigma model metric remains semidefinite positive in all cases, due the to positivity
properties of the conformal-frame  sigma model metric.

\end{abstract}

\vspace{24pt}
\end{titlepage}


\section{Introduction}
In this note we comment on some general properties of the sigma-model metric in $N=1$ supergravity coupled to 
chiral multiplets~\cite{cfgvp,bawi}.
In particular we discuss properties of the metric in the conformal and Einstein frames.
These frames 
are particular cases in the superconformal approach to supergravity~\cite{fvp}. Different frames are suitable for
uncovering the physics of different explicit models. Positivity properties of the sigma-model metric are maintained by the metric in different frames, but the sigma model metric in the conformal frame can have lower rank than in the 
Einstein frame,
 before elimination of the axial vector auxiliary field. In this note, we show examples where the conformal metric is non-invertible but the final sigma model metric is nevertheless positive definite. 

In the case of nonlinear realizations of supersymmetry, the nilpotency of the auxiliary field $A_\mu$ may reduce the rank of the full scalar metric. A particularly striking example is the inflaton, where the nilpotency of $A_\mu$ removes the 
pseudoscalar partner of the inflaton (the ``sinflaton''). 

\section{Supergravity in the Conformal and Einstein Frames}
Conformal-frame supergravity is the formulation that follows directly from the tensor calculus~\cite{fvn,sw,fvp} that uses a minimal set of auxiliary fields; namely,  a
complex scalar $u$ and and axial vector $A_\mu$. This corresponds to a particular choice of Jordan frame (see e.g.~\cite{kall}) in which the
frame function, $\phi(z,\bar{z})$, is the first component of a real superfield, $\Phi(Z,\bar{Z})$, whose local D-density is the 
Lagrangian of supergravity coupled to a nonlinear sigma model~\cite{cfgvp}
\beq
\Phi(Z,\bar{Z})\Big |_D= L_{CS}= e \Big[ {\phi \over 6} R - \phi_{i\bar{\jmath}}  \partial_\mu z^i \partial_\nu \bar{z}^{\bar{\jmath}}g^{\mu\nu} - {\phi\over 9} A_\mu A_\nu g^{\mu\nu} +{i\over 3}  A_\mu (\phi_i \partial_\nu z^i - \phi_{\bar{\imath}}
\partial_\nu \bar{z}^{\bar{\imath}} ) g^{\mu\nu}+... \Big].
\eeq{1}
Here $e=\det e^a_\mu$, $\phi_i=\partial_i \phi$ etc.\ and we only wrote explicitly bosonic terms relevant for our discussion. The function 
$\phi(z,\bar{z})$ is negative with non-negative sigma model metric $\phi_{i \bar{\jmath}}$. 
We call eq.~(\ref{1}) ``conformal frame Lagrangian'' because its Einstein equation is sourced by the improved 
energy-momentum tensor in curved space~\cite{ccj} (see \cite{vpr} for the supergravity extension). We remark that the conformal-frame Lagrangian is additive in the
$\phi$ function, differently from other Jordan frames. This property was important in formulating the  ``sequestering'' scenario of  
ref.~\cite{rs}. The physical properties of a supergravity  model may be transparent in one frame but hidden in another. 
For instance, the example of ref.~\cite{rs}, based on brane constructions, was naturally additive in $\phi$. 
The same is true for conformally coupled scalars. Other models, such as sequential toroidal compactifications 
of higher-dimensional supergravity, are instead additive in the K\"ahler potential $K$. 

The Einstein frame action is obtained by performing a Weyl rescaling of the vierbein,
$e^a_\mu \rightarrow e^a_\mu \exp(\sigma)$, 
 such that the $R$ curvature term coincides with the Einstein-Hilbert action.\footnote{In the superconformal approach, 
 different Jordan frames correspond to different gauge choices for the compensating multiplet~\cite{fvp}.} 
 This rescaling is
 \bea
 e^{2\sigma}&=&  -{3\over \phi}, \label{m2} \\
 {1\over 6} e \phi R &\rightarrow & -{1\over 2} eR -{3\over 4} e g^{\mu\nu} \partial_\mu \log \phi \partial_\nu \log \phi + 
 \mbox{ total derivative}.
 \eea{m2a}
Under this rescaling we have $L_{CS} \rightarrow L_{ES}$ with
\bea
L_{ES}/e &=&  -{1\over 2}R +{3\over \phi} \phi_{i\bar{\jmath}} \partial_\mu z^i \partial_\nu \bar{z}^{\bar{\jmath}} g^{\mu\nu}  
-{3\over 4} [(\log\phi)_i \partial_\mu z^i + (\log\phi)_{\bar{\imath} }\partial_\mu \bar{z}^{\bar{\imath}}]^2 + \nonumber \\ && 
+{1\over 3} A_\mu A_\nu g^{\mu\nu} 
-i  A_\mu [(\log\phi)_i \partial_\nu z^i - (\log\phi)_{\bar{\imath}} \partial_\nu \bar{z}^{\bar{\imath}}]g^{\mu\nu} .
\eea{m3}
Finally, if one integrates out the $A_\mu$ field, one gets 
\beq
L_{ES}/e =  -{1\over 2}R +{3\over \phi} \phi_{i\bar{\jmath}} \partial_\mu z^i \partial_\nu \bar{z}^{\bar{\jmath}} g^{\mu\nu}  
-{3\over 4} [(\log\phi)_i \partial_\mu z^i + (\log\phi)_{\bar{\imath} }\partial_\mu \bar{z}^{\bar{\imath}}]^2 
+{3\over 4} [(\log\phi)_i \partial_\mu z^i - (\log\phi)_{\bar{\imath}} \partial_\mu \bar{z}^{\bar{\imath}}]^2 .
\eeq{m4}
This is a nonlinear sigma model with K\"ahler metric ($d\equiv dz^i\partial_i$, $\bar{d}\equiv d\bar{z}^{\bar{\imath}}\partial_{\bar{\imath}}$)

\bea
(d \otimes \bar{d} ) K&=& -{ 3\over \phi} (d \otimes \bar{d})\phi  + {3\over 4} (d\log\phi + \bar{d}\log\phi)\otimes 
(d\log\phi + \bar{d}\log\phi) + \nonumber \\
&& -{3\over 4} (d\log\phi - \bar{d}\log\phi) \otimes (d\log\phi - \bar{d}\log\phi) \nonumber \\
&=& -{3\over \phi} (d \otimes \bar{d})\phi + 3 d \log \phi \otimes \bar{d} \log \phi = - 3 d \otimes \bar{d} \log \phi.
\eea{m5}

\section{Properties of the Sigma-Model Metric}
By inspection of eq.~(\ref{m5}), the Einstein-frame K\"ahler metric is the sum of three $2n\times 2n$ matrices:
the matrix $\phi_{i\bar{\jmath}}$ and two positive rank-one matrices, the first coming from  the Weyl rescaling and the second from
integrating out the $A_\mu$ field. The physical requirement is that $\phi_{i\bar{\jmath}}$ is non-negative and of rank 
$\geq 2n-2$. For $n=1$ the rank-zero example is the inflaton metric discussed in the next section. We also observe that
the splitting of the K\"ahler metric in three factors does not respect the K\"ahler invariance 
\beq
K \rightarrow K+ \Lambda 
+ \bar{\Lambda}, 
\eeq{m6}
where $\Lambda$ is a holomorphic function of the coordinates. In the $\phi$ variables, the transformation corresponds to 
\beq
\phi \rightarrow \phi e^{-{1\over 3} (\Lambda + \bar{\Lambda})}.
\eeq{m7}

Let us now consider a case where the $\phi_{i\bar{\jmath}}$ metric has rank $2n-2$. This is the model with 
\beq
\phi^3=-{1\over 3} d_{ijk} (z^i + \bar{z}^{\bar{\imath}}) (z^j + \bar{z}^{\bar{\jmath}}) (z^k + \bar{z}^{\bar{k}}) .
\eeq{m8}
This metric appears in the large-volume limit of the K\"ahler class moduli in Calabi-Yau compactifications of type IIA superstrings~\cite{cfg}. Since $\phi$ depends only on $\Re z^i$ and is homogeneous of degree one, it follows that 
$\phi_{i\bar{\jmath}}\Re z^j =0$. So $\phi_{i\bar{\jmath}}$ has a null eigenvector. Since the metric for $\Im z^i$ is the same as the one
for $\Re z^i$, it follows that the sigma-model metric splits into a rank-$2n-2$ part plus a rank-2 part
\beq
K_{i\bar{\jmath}}= -3\Big( {\phi_{i\bar{\jmath}}\over \phi} - {\phi_i \phi_{\bar{\jmath}} \over \phi^2} \Big).
\eeq{m9}
Notice that if we  only retain  the volume modulus, $t$, then $\phi_{t\bar{t}}=0$ and the entire metric resides in the 
last term, which, in this case, is the full metric.

We  remark that the conformal frame action is  invariant under the K\"ahler tranformation~(\ref{m6},\ref{m7}), even if the 
sigma model metric is not. This happens because the transformation 
also acts nontrivially on the conformal frame metric $g^C_{\mu\nu}$:
\beq
g_{\mu\nu}^C\rightarrow g_{\mu\nu}^C e^{{1\over 3}(\Lambda + \bar{\Lambda})}.
\eeq{m9a}

\section{Inflaton Disk Geometry}
In many supersymmetric models such as the supergravity extension~\cite{cfps,rr2} of the $R+R^2$ ``Starobinsky'' 
model~\cite{star1,star2}, the inflaton 
$\varphi$ has a K\"ahler potential  
\beq
K=  -3 \log [(\varphi+ \bar{\varphi})/3].
\eeq{m10}
The standard inflaton is the real part of $\varphi$ while the imaginary part is its supersymmetric partner, the ``sinflaton." 
As in the previous example, $\phi_{\varphi,\bar{\varphi}}=0$. This is due to the fact that $R+R^2$ supergravity is dual to
a standard (conformal frame) supergravity with Lagrangian
\beq
L_{CS} = e {\phi \over 6} R +.... , \qquad \phi= - (\varphi+ \bar{\varphi}).
\eeq{m11}

This formula shows that the K\"ahler metric of this model is entirely due to curved space effects, since the two degrees of
freedom $\Re \varphi, \Im \varphi$ acquire kinetic terms only though the Weyl rescaling and the 
$A_\mu$ field equation. Dropping the $A_\mu$ contributions one obtains the $R+R^2$, $N=0$ theory~\cite{whitt}.

\section{Nonlinear Realizations}
Nonlinear realizations of local supersymmetry have been widely discussed in the recent past. Beyond the Volkov-Akulov~\cite{va}
nilpotent chiral superfield $X$ ($X^2=0$)~\cite{f-ks}, other superfields can undergo nonlinear realizations if they satisfy nilpotency
conditions. In supergravity one can impose constraints which have no analog in rigid supersymmetry, since they create
nonlinear restrictions on the underlying local superspace geometry. In this case, the only constraints that do not affect
the gauge field sector are nilpotency constraints on the auxiliary fields $u$ and/or $A_\mu$. They have the 
form~\cite{cdafp}
\beq
X\bar{X} ({\cal R}-c)=0, \qquad X\bar{X} G_{\alpha\dot{\alpha}}=0.
\eeq{m12}
Here ${\cal R}$ is the scalar curvature chiral superfield and $G_{\alpha\dot{\alpha}}$ is the real 
Einstein superfield~\cite{bawe}. These
constraints imply that $u$ and $A_\mu$ are nilpotent. The first constraint will affect the scalar potential and the latter will
affect the kinetic terms.  Recently these constraints have been used in higher derivative supergravity~\cite{ffkl}. In this note we discuss only the properties of the sigma model metric, so we confine our discussion to the latter constraint. 

Nilpotency of $A_\mu$ implies that its contribution to the kinetic term in eqs.~(\ref{1},\ref{m4}) should be deleted. This procedure in some cases 
has the effect of removing the propagating degree of freedom associated to on of the rank-one matrix contributions to the sigma-model metric. In particular, in the inflaton model discussed above, it deletes the ``sinflaton'' degree of freedom.

In the multi-field case with $\phi$ potential given by eq.~(\ref{m8}), the full metric will have rank $2n-1$ so that one scalar
degree of freedom will always become non dynamical. It is important however to make sure  that the remaining degrees of freedom have positive metric. This follows from the positivity conditions on the $\phi_{i\bar{\jmath}}$ metric. 
We also observe that two $\phi$ potentials related by coordinate transformations may have a different number of 
propagating degrees of freedom once the nonlinear constraint on $A_\mu$ is imposed. This is due to the fact that when
the $A_\mu$ field becomes nilpotent, the underlying geometry is no longer K\"ahler. This is easily seen in the inflaton
example. When $\phi= -(\varphi + \bar{\varphi})$ the sinflaton does not propagate. The potential 
$\phi=\varphi\bar{\varphi}-1$, which corresponds to a conformally coupled complex scalar, is obtained from the former, 
up to a  K\"ahler transformation, by
the coordinate transformation  $\varphi \rightarrow (\varphi +i )/(\varphi -i)$. The latter potential is easily seen to give a 
sigma model where two scalars propagate, since it corresponds to a flat metric $\phi_{\varphi,\bar{\varphi}}=1$ in the conformal frame. 
 
 \subsection*{Acknowledgments} 
   S.F.\ was supported in part by the CERN
TH Department and by INFN (IS CSN4-GSS-PI). M.P.\ would like to thank the CERN TH Department for its kind hospitality during completion of this work. M.P.\ was supported in part by NSF grant PHY-1620039.

\end{document}